\documentclass[a4paper,11pt]{article}
\usepackage{amsmath}
\usepackage{amssymb}
\usepackage{amsfonts}
\usepackage{graphicx}
\usepackage{caption}
\usepackage{subcaption}
\usepackage[utf8]{inputenc}
\usepackage[T1]{fontenc}
\usepackage{authblk}

\begin{document}
\title{ Persistent order in Schramm-Loewner Evolution driven by the primes last digit sequence }
\author[1]{Theophanes Raptis}
\author[2]{ Alberto Fraile}
\affil[1]{Physical Chemistry Lab, Chemsitry Dept., University of Athens}
\affil[2]{Nuclear Futures Institute, Bangor University, Bangor, LL57 1UT, UK}
\maketitle

\abstract{We report on a peculiar effect regarding the use of the prime's last digit sequence which is equivalent to a quaternary symbolic sequence. 
This was used as a driving sequence for the recently introduced Schramm-Loewner Evolution after using two different classes of possible binary encodings. 
We report on a clear deviation from the standard space-filling curves normally expected from such a process. We also contrast this behavior with others 
produced via a symbolic dynamics applied on standard noise sources as well as deterministic sequences produced by simple automata. Our findings include 
the well known, Morse-Thue sequence as the simplest model exhibiting such behavior apart from some strongly biased Levy walks.  }

\section{Introduction}

The random walk idea is one of the most ubiquitous concepts of statistical physics and it lends applications to numerous scientific fields 
(see, e.g., \cite{Barber}-\cite{Hod} and references therein). An important class of random walks hes been recently associated with a special process that
became known as the  "Schramm-Loewner evolution" (SLE). Impressive progress has recently been made in the field of critical lattice models using the SLE theory. 
In SLE, random critical curves are parametrized by a single parameter k, related to the diffusivity of Brownian motion \cite{gamsa}.

The SLE were first introduced in \cite{sle} where they were termed "Stochastic Loewner Evolutions" due to much earlier work on the Loewner
equation \cite{loewner} who demonstrated that an arbitrary curve not crossing itself can be generated by a real function by means of a conformal transformation. 
A real 'driving' function defined in one spatial dimension (1D) thus encodes a curve in 2D, in itself an intriguing result. 
In 2000 Oded Schramm  extended this method and demonstrated that driving the Loewner evolution by a 1D Brownian motion, the curves in the complex plane 
become scale invariant; the fractal dimension turns out to be determined by the strength of the Brownian motion.
 
Due to the arbitrary character of the driving
function, an SLE process can be used to produce a very large variety of possible paths on the complex plane including both smooth and stochastic
ones. Some efficient algorithms for computing such paths have been prescribed by Kennedy in \cite{kennedy}.
Examples of systems described by SLE include familiar statistical models like the Ising Spin, it's generalization in the Potts model\cite{pot}, polymer theory\cite{poly1},\cite{poly2} 
as well as geometric critical phenomena like percolation, self-avoiding random walks, spanning trees and others. Many reviews of SLE and its applications in physics already exist, see 
for instance \cite{Bauer}-\cite{Kager}.

Our interest in this class of processes stems from the possibility of extending previous work due to the possibility of using a large binary stream as a driving
function. Specifically, it is interesting to associate such large bit streams with the distribution of prime numbers because of a peculiar effect first noticed by Chebyshev.
This originated in previous work \cite{fraile1}, \cite{fraile2} where a set of new geometrical methods were proposed for the study of the primes numbers distribution.
Apart from a variety of unsolved problems in pure mathematics involving primes, the prime number distribution is also fundamental in  many other areas of practical applications 
such as cryptography, internet security\cite{dicky}\cite{khairina} etc.

Regarding the later, an early observation by Chebyshev also verified by others \cite{Jerzy}\cite{Sounda} concerned a serious imbalance of digits pertaining to a congruence class
suggesting an underlying bias in the overall distribution of unknown origin. Moreover, previous examination of the so called last digits sequence resulting from taking their 
$(mod 10)$  congruences is  known to be equivalent to a quaternary symbolic sequence and has been shown in \cite{fraile2} to have some unusual characteristics when 
projecting it as a 2D random walk via a simple and intuitive binarization protocol. 

Due to the fact that such a low alphabet can be easily turned to a binary stream serving  as a driving for an SLE process it was both suggestive and  interesting to utilise the great sensitivity 
of such a process in order to further obtain numerical evidence for the presence of a possible structure in the prime distribution based on a direct comparison with "$1/f$" noise
 generators when used as alternative standard drivers. Such an attempt is in the same spirit with previous proposals on the use of symbolic dynamics methods for number
theoretic problems \cite{lacasa}

In this work, we report on a particular use of one such process where the driving function is given in terms of a symbolic sequence in a binary alphabet
 selecting between the two possible branches of one of the general solutions of the Loewner equation.  Two classes of such binary sequences are
contrasted, one being a deterministic set including the primes last digit and, in second place, some representatives of random ones. The later were obtained after applying 
symbolic dynamics on certain known noise sources given an adjustable threshold on their main random variable domain. 

In section 2, we briefly introduce the SLE algorithm and we comment on certain observations from numerical experiments. Specifically,we report on a peculiar 
extremal behavior presented by the last digit sequence when compared with other standard noise sources.
We also report on certain properties that appear persistent in such a case while, adjusting the symbolic dynamics threshold over the domain of some standard Levy walks 
reveals two possible extremes approachng either the real or the imaginary axis indefinitely. We also contrast these results with other known
deterministic sequences to find that the simplest model reproducing a similar behavior appears to be the well known, \textit{Thue-Morse} sequence. 

In section 3, we attempt to elucidate some of the peculiar characteristics found from the driven SLE based on certain statistical properties related to the
block length statistics as obtained via a run length analysis \cite{Pu} a well known compression method.

\section{ Schramm-Loewner evolution as a discriminant of sequences complexity }

\subsection{Preliminary definitions}
An SLE process is generally defined via the solutions of a stochastic differential equation which when defined on the upper half plane takes the form of
a Loewner stochastic process as

\begin{equation}
\partial_t f_t(z) = 2(f_t(z) - \xi(t))^{-1}
\end{equation}

with the additional condition $f(\gamma(t)) = \xi(t)$ where $\xi$ is a driving function taking values on the boundary and $\gamma$ 
a planar curve. Additionally, the solution is supposed to satisfy a "hydrodynamic normalization" condition as
$f(z)\rightarrow z + cz^{-1} + O(z^{-2}) $ as $z\rightarrow \infty$. We note that (1) can also adopt the form of a standard Langevin process via a change of
variables $w = f(z) - \xi$ as

\begin{equation}
\partial_t w_t(z) = 2w_t(z)^{-1} - \xi(t)
\end{equation}

There are two branches of such solutions given in the literature in the form

\begin{equation}
f_{\pm} = (z \mp a)^a(z \pm b)^b \\
\end{equation}

The exponents in (3) have a complicated dependence on a critical parameter given as $a =( 1 - \sqrt{k/(16 + \kappa)} )/2, b = 1 -a$
which can also be simplified as $ a = (1 - c)/2,  b = (1+ c)/2, c = 1/\sqrt{1 + 16/ \kappa}$ where $\kappa $ guides through different types of known 
processes like self-avoiding walks ($\kappa\leq 8$) while for values greater than 8 it produces space filling curves over the plane.

An SLE interface is produced by the iterative application of the two possible conformal maps leading to sequences of the form
$f_{X_1}\circ f_{X_2}\circ ... f_{X_n}$ where $X_n$ stands for some dichotomic random variable choosing between branches
and all paths start at $z=0$. Each path is formed by a linearly increasing sequence of words or functional composites as
$$
[ 0 f(0, X1)]
$$
$$
[ 0 f(0, X2) f( f(0, X1), X2)]
$$
$$
[0 f(0, X3) f( f(0, X2), X3)  f( f( f(0, X1), X2), X3)]
$$
$$
....
$$
each contributing a new point in the resulting curve.  The overall complexity of this scheme is similar to an arithmetic progression  scaling as $N(N+1)/2$.

 It should also be pointed out that similar curves can be obtained reading a sequence in reverse order which although irrelevant for certain random sequences 
it could influence some of the results from determinisitc ones like those tested in the second part of this work. 
Extensive testing with reversed sequences revealed no significant differences. To facilitate this iterative scheme it is possible to combine the two branches into a single 
expression utilizing the internal symmetry in the algebraic representation of $f_{\pm}$ in the form

\begin{equation}
f_{\sigma} = ( z - \sigma a )^a (z - \sigma a + \sigma)^{1-a}
\end{equation}

using $\sigma_n = 2X _n-1$  as the corresponding sign variable  projecting from a Boolean value of $X_n$ to the ${\pm 1}$ alphabet. 
An example Matlab code is already given in \cite{Xgit} which is easy to modify accordingly. In order to facilitate processing of sufficiently large samples of primes 
we had to enable the use of a compiled version via a complete transcription into an equivalent compiled fortran 95 code. The code is provided as supplementary material.

Although one can process millions of primes with this code, certain figures are based on
smaller samples to avoid visualization problems. We also provide a short tabulated form of various acronyms used throughout the text in Table 1 for ease of reference
since there are many different naming conventions involved for the complete characterization of each separate noise source and the encodings used.

We should stress the fact that the particular process is here used as a kind of 'black box' for performing unbiased numerical eperiments due to its possible sensitivity 
able of detecting and amplifying unusual statistics over large portions of symbolic sequences. 
Our purpose was solely to treat the SLE curves as a possible indicator of underlying pattern 
complexity of the driving bitstreams entering (3) via the indices defining a separate branch each time. 

The PLD sequence can also be found online in the well known Encyclopedia of Integer Sequences (OEIS) \cite{oeis}. The first four digits were always omitted
in our experiments since they are the only ones not conforming with the quaternary alphabet.
Due to the unusual and distinct character of the behavior observed in numerical experiments we also incorporated several known binary sequences that are
not random but they are produced by certain simple substitution rules or automata, for this reason being collectively called \textit{Automatic Sequences}. Some
of the most well known are the \textit{Thue-Morse} (TM), the \textit{Baum-Sweet} (BS) and the \textit{Rudin-Shapiro} (RS) sequences. A thorough analysis of the
 properties of such automatic sequences has been presented in a classical monography by Allouche \cite{auto}. A TM driven walk
shall be denoted as TMW and similarly for the rest of the automatice sequences.

In principle, one could merely run the same process over the 
global set of all $2^L$ possible bit strings although this is beyond our capacity. On the other hand, there is a crucial difference between the PLD
sequence as well as some automatic ones,  and those produced by arbitrary noise sources. 
In particular, both the PLD and the automatic ones are fixed, determinisitc sequences based on some underlying mathematical property, 
either that of divisibility or some substitution automaton. As such, they thus belonging to a very "thin" subset in comparison to those sampled by other noise sources.

\begin{table}
  \centering
\begin{tabular}{ |c|c| } 
\hline
Abbreviation & Meaning\\
 \hline
 PLD &   Primes Last Digit  \\ 
 RPLD & Reduced PLD  \\ 
 TLW &  Thresholded Levy Walk  \\
 MFW & Mittag-Lefler Walk \\
 TMW & Thue-Morse driven Walk \\ 
 \hline
\end{tabular}
\caption{Acronym conventions used throughout this document.}
\end{table}
 
\subsection{Comparisons based on numerical experimentation}

We contrasted the behavior of the resulting curves when driven from different noise sources versus the one given by the PLD sequence. 
The various noise generators are treated as arbitrary dynamical systems extracting from each a particular symbolic dynamics using 
a thresholding scheme via the assignment $X_n \leftarrow r_n > t$ for some adaptable threshold $t$ value. 

The types of noise sources used for comparison included a standard flat  and a normal random generator with $t = 1/2$ and $0$ respectively,  
corresponding to unbiased, equiprobable density binary sequences as well as certain coloured noise sources including  certain $1/f$ noise sources 
also known as "pink" or "flicker" noise. 

We also tested  a Thresholded Levy Walk (TLW) with a variable threshold taken in $[t, \infty ]$ for some small $t > 0$. 
A standard implementation of a noise source for both white,  Brownian and pink noise exists in the harmonic analysis package LTFAT\cite{LTFAT}.

Additionally we tried a special noise generator known as the "Mittag-Lefler noise"\cite{MLnoise}. This comes from a special transformation of a 
uniform random variable leading to distributions that are non-singular near the origin while still exhibiting the $1/f$ fat tails asymptotically.

In the case of the PLD sequence, there are two alternative possibilities for translating from the original quaternary alphabet of the $\{1,3,7,9\}$
symbols to a pure binary one, always omitting the first four primes. 
We may classify them as a \textit{"lossless"} one giving rise to the PLD case and, a \textit{"lossy"} one which we choose
referring to as the Reduced PLD (RPLD) since it always results in half the length of the PLD case.  These terms do not imply any sort of compression
mechanism. The two different protocols are as below
\begin{itemize}
\item The lossless case is based on a direct substitution of values in the set $\{1, 3, 7, 9 \}$ with the two block strings $\{ 00, 10, 01, 11\}$ as a continuous
      	 concatenation of two symbol blocks. 
\item The lossy method makes a direct correspondence of any of the  $2^4 - 2 $ possible binary patterns by  assigning a single bit in every symbol as in a logical table
	 excluding the all zeros and all ones patterns. These cases are shown in Table 2  using lexicographic ordering. Any such table contains different groups characterized by 
	their \textit{weight} defined as the summand of one bits. This allows grouping different combinations by the associated binomial coefficient $C(4, i), i=1,..,3$. 
        There can only be three main groups possible of a total of the 14 possible patterns with weights from 1 to 3.  Two of them contains all bit shifts (weigtht 1) and their 
	not-complements (weight 3) while the central group of weight 2 with $C(4, 2) = 6$ is the only non biased one since there is always equal number of zeros and ones.  	
	 For instance , an equipartitioned code for PLD binarization corresponds in taking $\{ 1, 3\}\rightarrow 0$ and $\{7,9\}\rightarrow 1$  or the $0 0 1 1$ code according
	to Table 2 while another would be given by the substitution $\{ 1, 7\}\rightarrow 0$ and $\{3,9\}\rightarrow 1$ or the $0 1 0 1$ code.    
\end{itemize}
The main reason for testing also the second case of the RPLD is the fact that for the lossy case, there is no particular prior reason for favoring any of the totality of possible
encodings. Actual experimentation showed that there is an important division between the unbiased group of weight 2 with the rest although all groups are associated with
some extremal behavior. The main observations can be summarized as follows

\begin{table}
  \centering
\begin{tabular}{ |c|c|c|c|c|c| } 
 \hline
Index & 1 & 3 & 7 & 9  & Weight\\
 \hline
 1 &1 &   0 & 0 & 0  & 1\\ 
 2 & 0 & 1 & 0 & 0 & 1\\ 
3 & 1 & 1 & 0 & 0 & 2\\
 ..& .. & .. & .. & .. &..   \\
12 &  0 & 0 & 1 & 1 & 2\\
13 & 1 & 0 & 1 & 1  & 3\\ 
14 &  0 & 1 & 1 & 1  & 3\\
 \hline
\end{tabular}
\caption{Lexicographic arrangement of alternative binarization schemes for the RPLD sequence.}
\end{table}

Results from all numerical experiments are summarized in Table 3, while a more detailed explanation is given as follows, including several graphical depictions of our
main findings. The 'Standard' characterization used in the second column in Table 3 is just a shortcut for the normally expected behavior of a generic SLE process as
found in the relevant literature followed by the usual form of phase transitions driven by the $k$ parameter.
\begin{itemize}
  \item The standard flat and Gaussian sources used as a reference always give the expected types of random curves resembling random walks on the complex plane
            as well as the standard phase transitions expected from different values of the driving parameter $k\in[0,8]$. 
	    This is also reflected in fig. 1(b) with respect to a critical behavior of both the PLD/RPLD cases as explained below.
        
   \item The RPLD case behaves in a more erratic manner with an increasing value of the $\kappa$ parameter while the simple PLD effect persists even beyond $\kappa = 8$
		where one expects a space filling curve.       
  
  \item The case of both the PLD and RPLD sequences is separate being uniquely characterized by two different asymptotic conditions given roughly as $Im(z) >> Re(z)$ 
 	   for the non-biased group of codes (see Table 2) with weight 2 and its complement $Im(z) << Re(z)$ for all other groups of codes.
           
   \item In the first case, the imaginary part always evolves  approximately like $a\sqrt{n}$ for some $a < 1$ where $n$ the iteration index for each new point. A standard fitting
	    routine showed this to be relatively stable with $k$ (see fig. 2) with an exponent close to $1/2 + \epsilon, \epsilon<<1$.
                      
   \item Standard $1/f$ noise sources also exhibit similar  behavior with  $Im(z) << Re(z)$ growing fast towards either the positive or the negative real axis. 
	 The case of a Levy walk (TLW) has a additional sensitive dependence  on the value of the threshold parameter $t$. 
	 As the parameter varies towards zero it also falls near the real axis while moving  beyond some value   near  0.2 the resulting curves tend to follow 
       almost parallel to the imaginary axis.
           
   \item A Mittag-Lefler (non-singular) noise generator does not reproduce the previous extermal behavior and behaves more like normal random sources.

  \item When the Thue-Morse sequence is used as driving for an associated random walk(TMW) also reproduces the $Im(z) >> Re(z)$ behavior. No other
     	    standard automatic sequence appears to have a similar effect.
\end{itemize}

\begin{table}
  \centering
\begin{tabular}{ |p{4cm}|p{4cm}|p{4cm}|  }
\hline
\multicolumn{3}{|c|}{SLE Classes} \\
\hline
Driving source & Curve Types & Parameter Sensitivity\\
\hline
Flat Distr. & Standard & Phase transitions \\
Gaussian Distr. & Standard & Phase transitions \\
PLD/RPLD  (w=2) & Im(z) >> Re(z) &  Indifferent \\
PLD/RPLD (w=1,3) & Im(z) << Re(z) &  Indifferent \\
TLW (thresh.<<1)     & Im(z) << Re(z) &  Indifferent \\
MLW & Standard & Phase transitions. \\
TMW &  Im(z) >> Re(z) &  Indifferent \\
\hline
\end{tabular}
\caption{Observed behavior for different classes of noise sources}
\end{table}

In figure 1(a) we show an example of a 1000000 points SLE created from an equal number of primes using the RPLD sequence, showing the SLE curve growing rapidly along
 the imaginary axis.  In figure 1(b) we show an overlap of the imaginary parts for all SLEs  where   the maximal blue curve corresponding to the RPLD sequence which exhibit the 
smoothest  possible evolution with respect to all the others.  We only show a small sample for convenience of visualization but the reader can readily explore this behavior for 
larger samples with the code provided in the supplementary material section.

\begin{figure}[!tbp]
  \begin{subfigure}[b]{0.5\textwidth}
    \includegraphics[width=\textwidth]{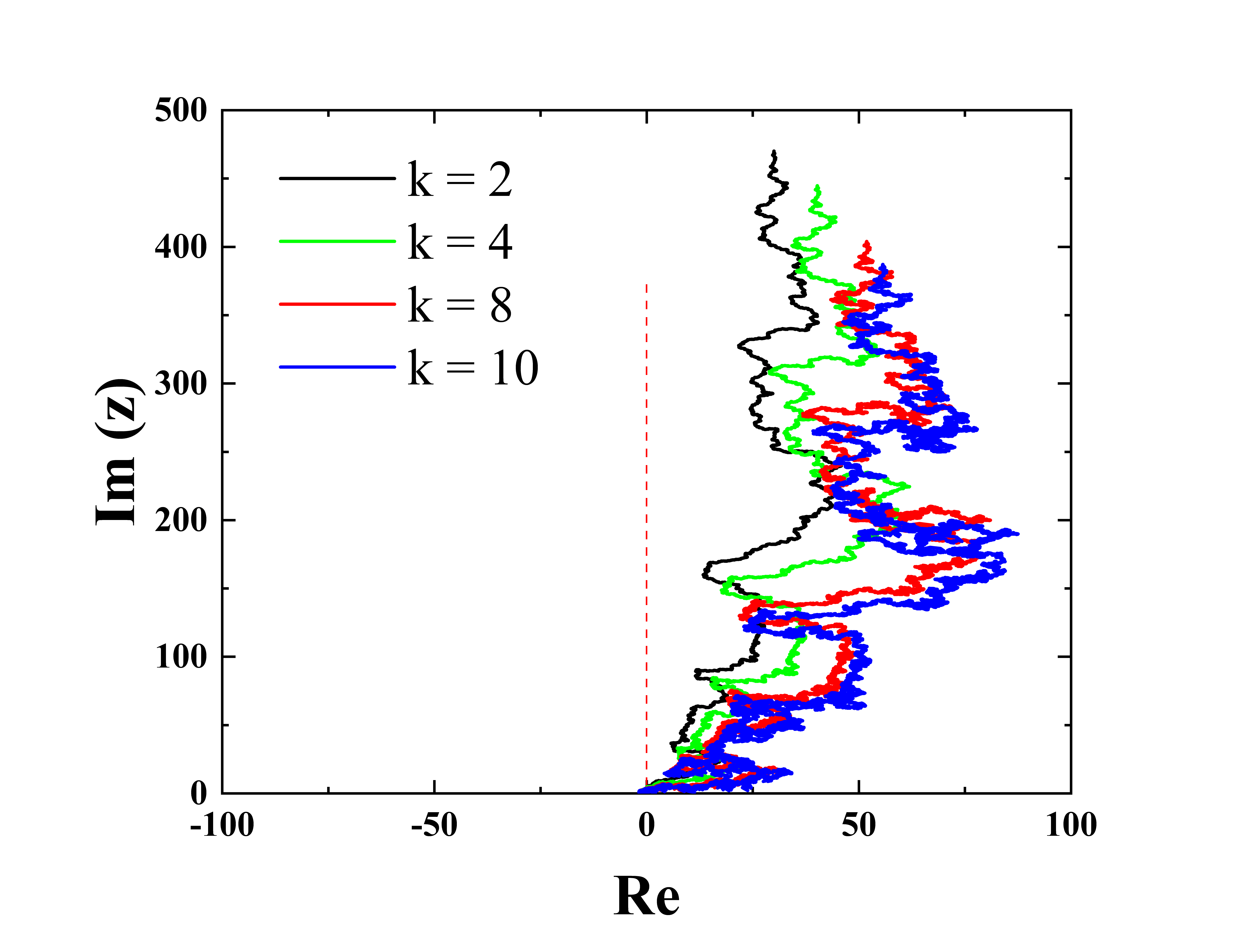}
     \label{fig:f1}
  \end{subfigure}
  \hfill
  \begin{subfigure}[b]{0.5\textwidth}
    \includegraphics[width=\textwidth]{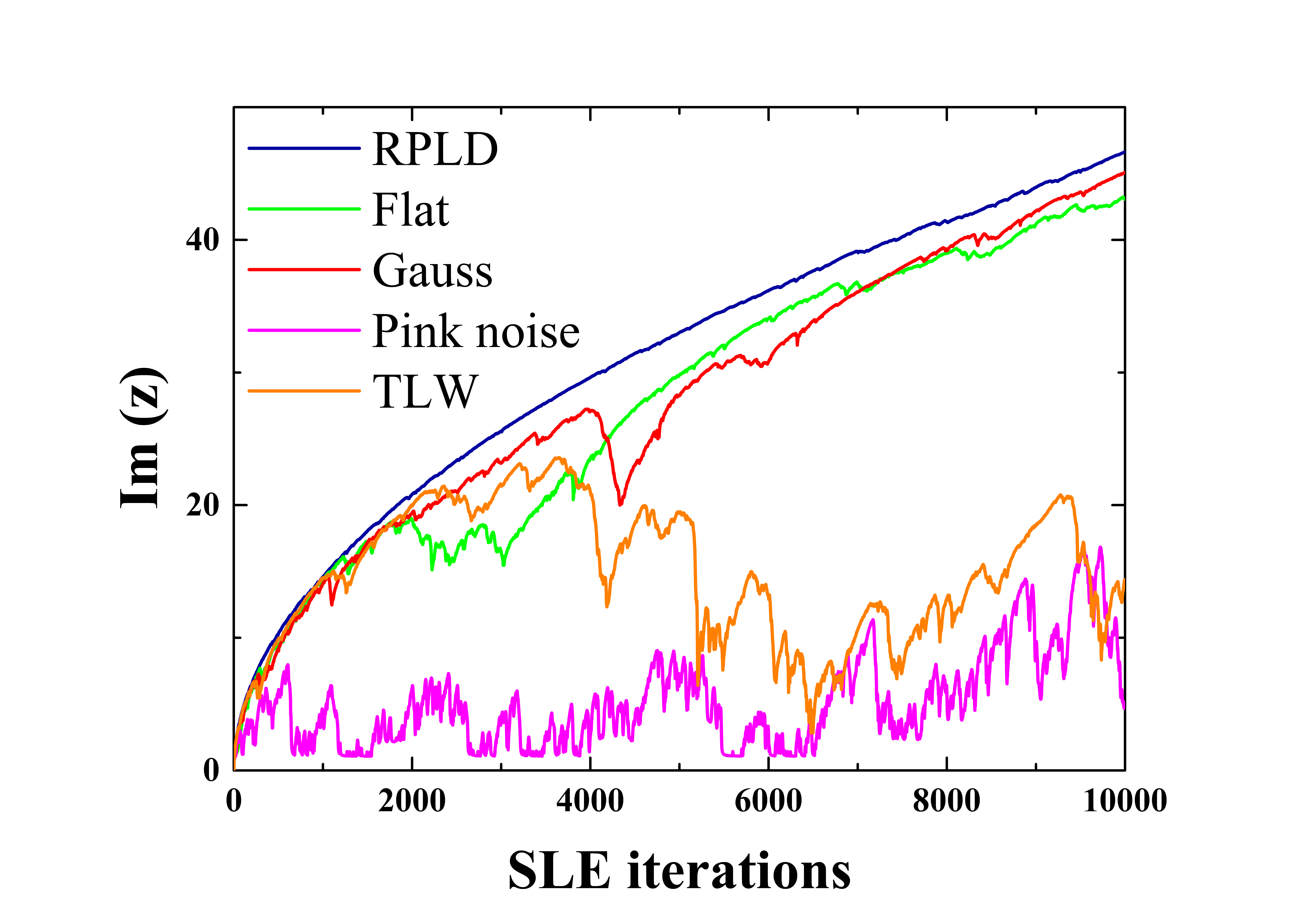}
    \label{fig:f2}
  \end{subfigure}
  \caption{ (a)First million points of an SLE curve for the PLD sequence and (b) overlap of all SLE imaginary parts for different noise sources.}
\end{figure}

In fig. 2(a), a plot of different RPLD with the iteration index shows the rather clear $\sqrt{N}$ persistence. 
In fig. 2(b) we also show a plot of different fits of the form $y=ax^b$for the exponent in the dependence of $Im(z)$ with $\kappa$.

\begin{figure}[!tbp]
  \begin{subfigure}[b]{0.5\textwidth}
  \includegraphics[width=\textwidth]{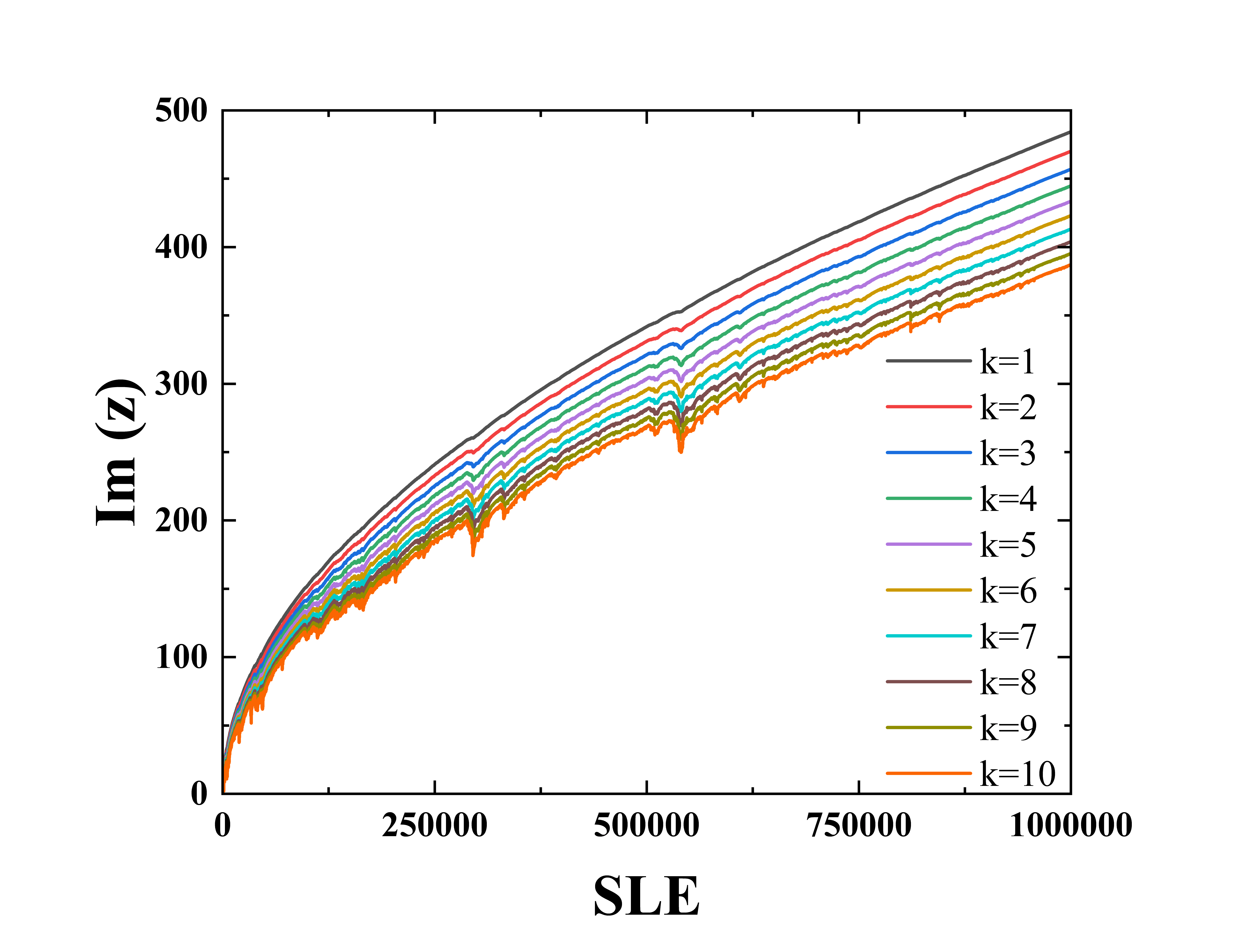}
  \label{fig:f1}
  \end{subfigure}
  \hfill
  \begin{subfigure}[b]{0.5\textwidth}
    \includegraphics[width=\textwidth]{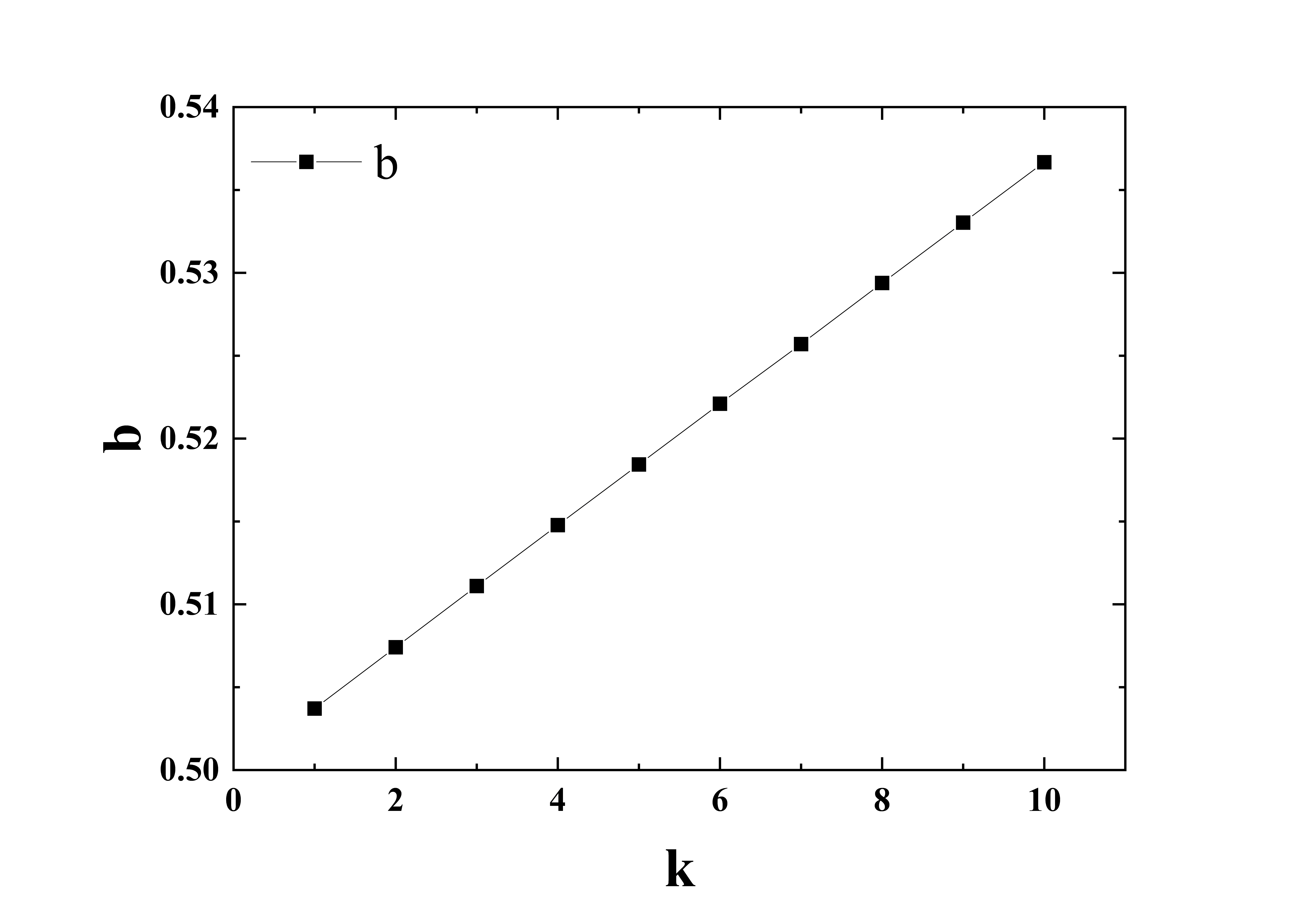}
    \label{fig:f2}
  \end{subfigure}
  \caption{  (a) First million points of SLE with RPLD and (b) exponent fits with $\kappa$ of the ten curves in 2(a).}
\end{figure}

In fig. 3, we also show the same behavior out of the Thue-Morse driven walk (TMW) where the same extremal behavior appears much more clearly pronounced.

\begin{figure}
  \centering
  \includegraphics[width=10cm]{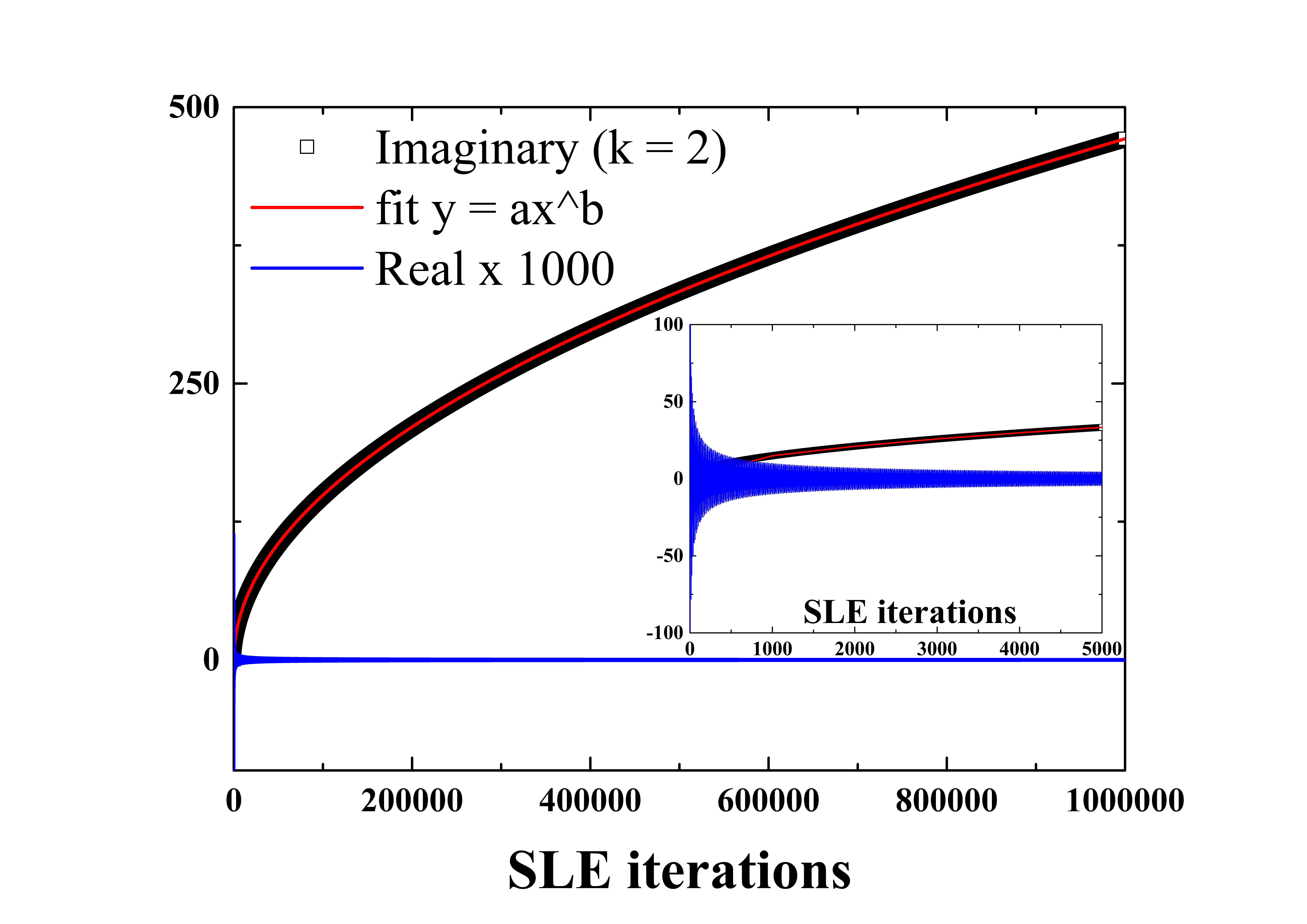}
  \caption{Real and imaginary parts of the SLE curve driven by the Morse-Thue sequence }
  \label{Fig 3}
\end{figure}

The case of TLW offers more intermediate possibilities due to the potential of lowering the threshold asymptotically towards zero thus effecting
also a drop of the overall curve obtained inside a small strip above the real axis. Hence, the particular class of TLW appears to be able to cover
a broad spectrum of behaviors between the two extremes found in the previous section. 

Motivated by this we also examined an increasing bias on
other white, normal and Brownian noise generators by moving the threshold values appropriately the result being always one of the two
extreme behaviors restricting the curves obtained near the real axis. This leads deducing the existence of some rare, "thin" subset of possible
binary strings among the plethora of $2^N$ total for $N >> 1$ of whose the specific block structures reflect the particular properties to which an SLE
process responds similarly with the PLD/RPLD effect. 

To gain some more insight, we also used the Mittag-Lefler noise generator which is easily produced via a special transformation of a Gaussian random variable. 
Indeed in such a case, the extremalities of the standard TLW case disappear, allowing us to conjecture that what the SLE effectively classifies is possibly the presence 
of a singular distribution of the sampled noise source via its symbolic dynamics. The two remaining questions from the previous section concern the appearence of the
persistent $\sqrt{N}$ behavior of the imaginary part and the deeper connection between deterministic and sampled noisy sequences.

We used some of the most known 
examples, including the Thue-Morse (TM), the Baum-Sweet (BS) and the Rudin-Shapiro (RS) sequences for which we know the existence of simple production rules
thus being of lesser complexity than the PLD/RPLD case. Surprisingly, all three types of behavior are then
being observed. Specifically, the RS sequence is the one exhibiting the space filling character expected from high values of the $\kappa$ parameter, while 
the BM and the TM ones, correspond to the opposite sides of the spectrum with the later exhibiting the most smooth evolution of the imaginary part close
to the $\sqrt(n)$ curve while the real part exhibits a kind of dissipative oscillation, asymptotically approaching zero (see figure 3).

To check the possibility of a hidden symmetry between $1/f$ processes and the
deterministic RPLD sequence, we also checked a possible inversion under axes reversal which corresponds to the involution 
$\hat{L}(z) = \mathbf{i}z^*, \hat{L}^2 = Id$ where star denotes transposition, effecting an exchange of real and imaginary parts.
While this is not a symmetry of the original solution (3), our purpose was to check whether mixing of terms could result in an asynmptotic condition
like $|f( \hat{L}(z) ) - \hat{L}( f(z) )| < \epsilon$ for some small $\epsilon$.  Results were negative for all noise sources and arbitrary values of the $\kappa$ parameter.
We continue with some more tests on the possible origin of the extremal effects observed.

\section{ Remaining questions on the origin of the extremalities }

From the previous phenomenological evidence, there is at least one main result, concerning the formal coincidence between the deterministic PLD/RPLD
sequences and the much simpler automatic TM one. While we do not currently have an analytical explanation of where the persistence of the PLD/RPLD sequences 
effect originates, we can employ some more tools from the harmonic analysis of noise sources. 

One very useful tool is given in context of so called, \textit{Multiresolution Signal Analysis} is the spectrogram, a version of short-time Fourier transform \cite{stft} 
or windowed Fourier transform running over small portions of the spectrum each time. This allows for simultaneous time-frequency analysis revealing possible modulations 
underlying noisy signals. The output is then represented by a whole array containing both axes with the additional time axis allowing visualization of possible frequency shifts 
and other variations while computing  instantaneous spectra in an easy graphical manner.  

We made a direct 
comparison of the spectrograms using LTFAT's  facilities.  Figures 4(a - b) shows two samples, the first being a typical "white" noise source 
and the second a standard RPLD with weight 4 encoding, exhibiting a "lighter" colormap than a standard flat distribution thus implying an overall modulation 
over the spectral range. Next, we search for the possible origin of this effect in the overall block length statistics, usually associated with the run length analysis 
of symbolic sequences.

\begin{figure}[!tbp]
  \begin{subfigure}[b]{0.5\textwidth}
    \includegraphics[width=\textwidth]{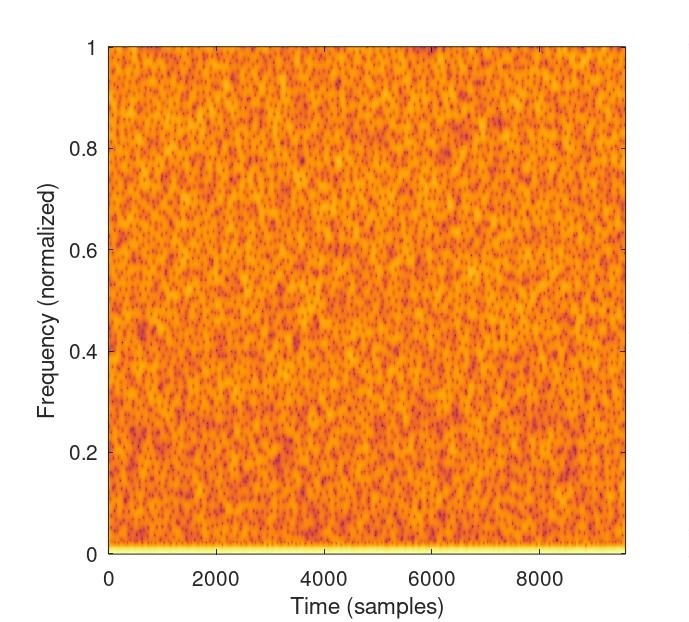}
    \caption{ }
    \label{fig 4}
  \end{subfigure}
  \hfill
  \begin{subfigure}[b]{0.5\textwidth}
    \includegraphics[width=\textwidth]{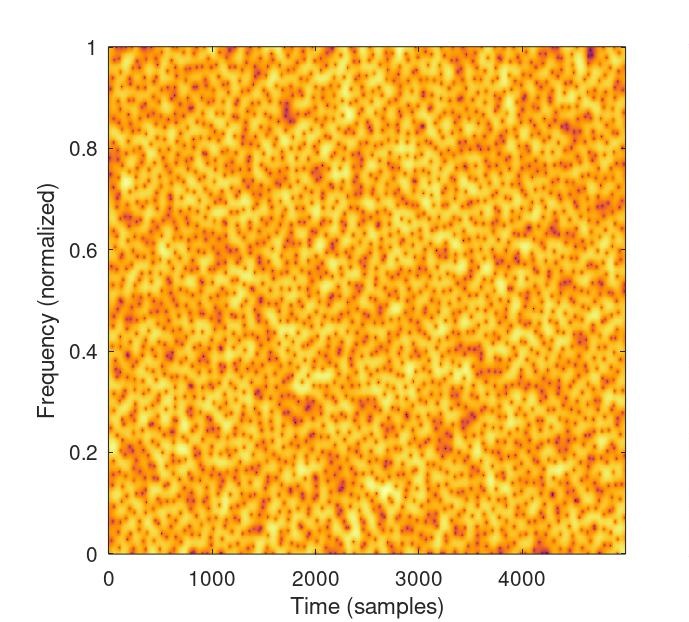}
    \caption{ }
    \label{fig 4}
  \end{subfigure}
  \caption{ Spectrograms for (a) white noise sequence and (b) the RPLD sequence.}
\end{figure}

Another way to discriminate between sequences with different pattern structures is given using an analysis of its run length encoding, a well known
compression method based on symbol block counting, by measuring the block lengths of consecutive one and zero blocks or alternating $\{\pm 1\}$ symbols. 
This always results in another alternating sign sequence of higher integer values for any binary input sequence. Making a precise integer histogram of 
the results for a large sample of both the RPLD with 78.500 primes and other noise sources including the TLW for threshold values adapted in the two extreme classes shows 
similar statistics as in figure 4, apart from two excess values due to proliferation of single symbol blocks.

\begin{figure}
  \centering
  \includegraphics[width=10cm]{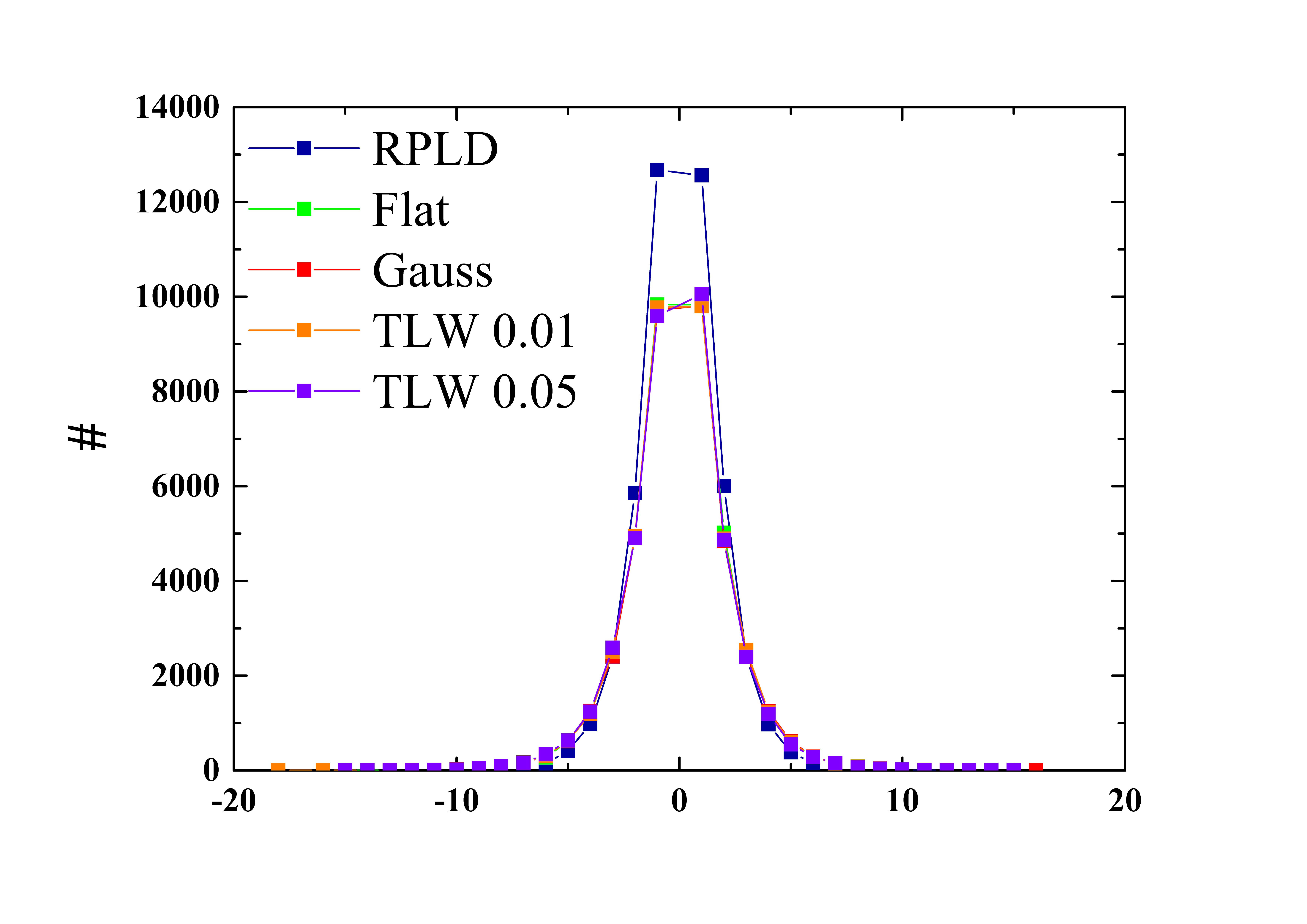}
  \caption{Comparative frequency histograms for zero (negative axis) or one (positive axis) block lengths from the symbolic dynamics of various noise sources vs the RPLD sequence. }
  \label{Fig 5}
\end{figure}

We should stress for the reader that the apparent appearence of a power law in figure 5, is not related to some so called, "Zipf's Law" \cite{zipf} which associates
frequencies with sub-word lengths. Instead, the frequencies appearing there denote repetitiveness of same symbols. In particular, the negative  $X$ axis
was used for the frequencies of same "zero" blocks and the positive axis for same "ones" blocks. 

The overabundance of single symbol blocks for the
RPLD sequence, suggests the origin of the extremalities observed to be related with noise sources that are singular near the origin. One can readily
verify that any constructible sequence with large blocks of same symbols leaving the trajectory on the same branch tend to create curves that slowly
fill the plane while, deliberately constructed sequences with rapidly changing small blocks prevent the space filling character of the resulting curves.

The run length statistics of the TM sequence is entirely different than that of all previous cases with only blocks of single and double symbols, either ones or zeros and with a perfectly symmetric histogram. On the other hand, it may allow a different interpretation for the origin of the RPLD persistence effect as a superposition of two processces, one automatic and the other stochastic. 

We note in passing that recent intensive
research by Ninagawa suggesting the association of $1/f$ statistics and computational universality or "Turing-completeness" were presented in \cite{shige}, \cite{nina}. 
Since several different types of universal automata appear to exhibit similar behavior one conjectures that the general class of such noise sources may
include Turing complete mechanisms as a special subclass. This in turn implies the possibility that deterministic sequences like the PLD/RPLD may hide a special 
hierarchy of subwords corresponding to some such hidden automaton associated at least with large fragments of this sequence spread out. This idea of coexistence is also
inspired by recent work \cite{noise} where concrete proofs have been given regarding the possibility of turning in principle uncomputable properties into computable
ones in the presence of injected noise. 
 
At the moment we can only conjecture that the TM case represents a basis for comparison which appears stable due to its independence from the diffusion driving 
$\kappa$ parameter, in contrast with the PLD or RPLD dependence shown in figures 2(a-b). There must then exist a contribution of unknown origin setting up this
instability.  

To further deepening our understanding of this peculiar effect we shall need the abstract definition of the MT sequence given in terms of one of the most fundamental
integer sequences characterising the binary representations of the integers with well known fractal characteristics, the so called, "Digit-Sum" \cite{ds} or
Hamming weights sequence with lots of applications in combinatorics and computer science \cite{comsci1}. 

Given a polynomial representation of every natural number as $\nu = \sum_i\sigma_i 2^i, \sigma\in\{0,1\}$ we can write this sequence with the aid of a local bit value map as

\begin{equation}
  w_H(\nu) = \sum_i^{L(\nu)} \sigma_i(\nu), \sigma_i(\nu) = \left \lfloor  \frac{\nu}{2^i} \right \rfloor(mod 2 )
\end{equation}

where $L = \left \lfloor log_2(\nu) + 1 \right \rfloor$ the maximal power of 2 present . The bit values given as $\sigma_i(\nu)$ represent a non-recursive decoding 
scheme based on shifting each integer backwards via integer division and extracting each bit at subsequent levels.

Given this sequence, the MT sequence is directly obtainable as 

$$(-)^{w_H(\nu)}\cong exp(\mathbf{i}\pi w_H(\nu)) \cong mod( w_H(\nu),2)$$

where the first two terms correspond to a $\{\pm 1 \}$ alphabet and the last to the standard binary one. 
As a matter of fact, the later is also identical with the binary parity of bit strings in 
common use for computer science applications eg, in error correction \cite{comsci2}. 

In previous work \cite{viral}, it was pointed out by one of the authors that many sequences computed by simple automata acting on a lexicographically ordered powerset of all strings of constant legnth, also termed a "Hamming Space", may inherit an internal self-similarity out of the periodicities involved  which can be seen as a type of semi-direct
 product  between a set of dilations and a set of automorphisms of the $\mathbb{Z}_2$ group.

This then allows an arithmetized form of any such based on an iterated list concatenation system utilising much simpler "reproducing maps" of lower complexity in the form

\begin{equation}
\mathcal{L}_n \leftarrow [L_{n-1}, \mathcal{M}(\mathcal{L}_{n-1})]
\end{equation}

The correspondence between (5) and (6) reduces to the choices $\mathcal{M}(x) = x + 1, \mathcal{L}_0 = [0]$ where $\mathcal{M}(x)$ is to be applied pointwise to every element of the previous list. 
For the associated MT sequence in the $\{\pm 1\}$ alphabet one simply has to use $\mathcal{M}(x) = - x, \mathcal{L}_0={-1}$.  Various other well known self-similar
sequences that admit similar generating maps are summarised in Table 4.

 \begin{table}
		\centering
		\begin{tabular}{|c|c|c|}
			\hline
			Sequence & $\mathcal{M}(x)$ & $\mathcal{L}_0$ \\
			\hline 
			Digit Sum ($h_w$) & $x_n+1$ & [0] \\
			Runlength terms     &  $(\hat{R})(x_n)+1$ & [0]\\
			Symmetric Gray   &  $(\hat{R})(x_n) + 2^{n-1}$ &  [0 1]\\
			Stern-Brocot	  & $(\hat{R})(x_n) + x_n$ & [1 2]  \\
			Gould   & $2x_n$   & [1]  \\
			\hline
		\end{tabular}
		\caption{ Reproducing maps for five known self-similar sequences ($\hat{R}(x_n) = x_{N-n}$ stands for a left-right reflection of a list with $N$ elements.) }
\end{table}

Based on this one can write any formal sequence of conformal map compositions as

$$
f_{\sigma_0}\circ f_{\mathcal{M}(\sigma_0)}\circ f_{\mathcal{M}(\sigma_0)}\circ f_{\mathcal{M}^2(\sigma_0)}\circ...
$$ 

Simple inspection of the list concatenation scheme in (6) shows that any symbol sequence follows as $\mathcal{M}^{w_H(i)}$ where $i$ the composites enumeration index.
Since the action of $\mathcal{M}$ for the MT sequence is identical with that of the $\mathbb{Z}_2$ group, in the above we have a case of a group modulation similar to the case 
of a semi-direct product between the set of automorphisms of $\mathbb{Z}_2$ and the discrete conformal group \cite{ssg}. 

The fact that this type of compositions exhibits a characteristic stability with respect to the SLE responce provides a possible start in understanding the kind of perturbations
responsible for the deviations taking place in the PLD or RPLD cases. Furthermore, there is a way to derive the TM sequence from a particular set of Fourier modes based
on the so called, \textit{Rademcaher System}\cite{Rad}, also termed a "lacunary" sequence due to the exponential gaps in its spectrum. 

 Given a sampling operator $\hat{\mathbb{S}}$ over an harmonic function such that the sign is interpreted as a unique symbol for each half-period, one can also write

\begin{equation}
  w_H(\nu) = \sum_i^{L(\nu)} \hat{\mathbb{S}}( sin( \pi 2^i t  ) )
\end{equation}

We can then associate the above with a corresponding ODE system of independent oscillators $\ddot{x}_i = -(\omega_i^2/2) x_i$ with $\omega_i =  2^i\pi $.

This allows conjecturing the existence of a possible class of stochastic dynamical systems giving rise to some similar behavior, intermediate to that between the TMW 
and the PLD and RPLD cases. Such could be produced from the symbolic dynamics of the collective variable $<\mathbb{S}(y_i)>(mod 2)$ from the activity of a
high dimensional overdampled Langevin dynamics\cite{Lan} with some  noise source of appropriate spectral characteristics in the form

\begin{eqnarray}
  dx_i & = & p_i dt \\
  d p_i & =  &  -(\omega_i^2/2) x_i dt  + \sqrt{\kappa} \sigma_i(t)
\end{eqnarray}

The problem then is to find sets of noise sources able to reproduce the persistent effects when their symbolic dynamics is used as a driving sequence for an SLE process.
This much more general problem lies beyond the scope of the present short report and is deferred for future examination.

\section{ Discussion and Conclusions }

Motivated by recent research in properties of the primes "last digit" sequence (PLD), we proceeded exploring its unusual properties based on the idea of using different
dynamical systems as possible classifiers of complexity when driven by such a sequence. The case of the Schramm-Loewner evolution (SLE) was chosen based on both
the possibility of driving via a dichotomic noise source and its sensitivity in any external modulating signal manifested in the geometric characteristics of the resulting
path. 

We uncovered an unexpected type of behavior where the deterministic PLD randomness or some strongly biased Levy walks show a persistent tendency of 
aligning with one of the two axes in the upper half plane. A most striking characteristic was the appearence of a $\sqrt{n}$ curve for the imaginary parts in all cases
where alingment was towards the imaginary axis. We also contrasted this behavior with other simpler deterministic sequences of lesser complexity which revealed that
the so called, \textit{Thue-Morse} (MT) sequence may serve as a model for this type of behavior with the most clear sign of the square root shaped curve for the imaginary
part.

We should stress the fact that our findings are only indicative at least regarding such deterministic sequences like the PLD or RPLD, since we have only applied this method 
for finite samples in the absence of any formal proof. It is then still possible that such behavior could be just a very large transient. 
This though, may not hold for the MT model sequence since its structure is much simpler based
on an internal set of symmetries reflecting the associated sum-of-digits sequence. This is also known to be associated with an increasing set of exponential intervals as 
$\left [0,..., 2^n - 1\right]$ mediated by a ssimple generating map, such that the resulting type of modulation over the SLE should not be expected to show any significant 
qualitative differences over increasing sample lengths. 

The similarity of the observed persistent effect which strongly deviates from standard space filling curves usually met in SLEs was only observed for the two cases of deterministic
sequnces, the PLD and the MT cases which does not allow to fully explore the nature of the special subset of sequences sharing similar properties. The only other case of strongly 
biased Levy walks was compared with other types of $1/f$ noise sources for which this effect fails allowing us to conjecture that the main characteristic associated with this
effect may also have to do with the singular nature of their distributions near zero.

\end{document}